\documentclass[manuscript]{aastex}
\usepackage{epsfig}
\shorttitle{Atlas of Vega} \shortauthors{H.-S. Kim et al.}
\begin{document}
\title{Atlas of Vega: 3850 -- 6860 \AA
\thanks{Based on data collected with the 1.8-m telescope operated at BOAO Observatory, Korea}
}

\author{Hyun-Sook Kim}
\affil{Department of Astronomy and Atmospheric Sciences, Kyungpook
National University, Daegu, Rep. of Korea}
\email{hskim503@knu.ac.kr}

\author{Inwoo Han}
\affil{Korea Astronomy and Space Sciences Institute, 61-1,
Hwaam-dong, Yuseong-gu, 305-348, Daejeon, Rep. of Korea}
\email{iwhan@kasi.re.kr}

\author{G. Valyavin}
\affil{Observatorio Astron\'{o}mico Nacional SPM, Instituto de
Astronom\'{i}a, Universidad Nacional Aut\'{o}noma de M\'{e}xico,
Ensenada, BC, M\'{e}xico} \email{gvalyavin@gmail.com}

\author{Byeong-Cheol Lee}
\affil{Korea Astronomy and Space Sciences Institute, 61-1,
Hwaam-dong, Yuseong-gu, 305-348, Daejeon, Rep. of Korea}
\email{bclee@boao.re.kr}

\author{V. Shimansky}
\affil{Kazan State University, Kremlevskaja Str., 18, 420008 Kazan, Russia}
\email{Slava.Shimansky@ksu.ru }

\and

\author{G.A. Galazutdinov}
\affil{Department of Physics and Astronomy, Seoul
National University, Gwanak-gu, Seoul 151-747 Rep. of Korea}
\email{runizag@gmail.com}

\begin{abstract}
We present a high resolving power ($\lambda$/$\Delta\lambda$ =
90,000) and high signal-to-noise ratio ($\sim$700) spectral atlas
of Vega covering the 3850 -- 6860 \AA\ wavelength range. The atlas
is a result of averaging of spectra recorded with the aid of the
echelle spectrograph BOES fed by the 1.8-m telescope at Bohyunsan
observatory (Korea). The atlas is provided only in
machine-readable form (electronic data file) and will be available
in the SIMBAD database upon publication.
\end{abstract}


\section{Introduction}

 Vega($\alpha$ Lyr = HD 172167 = HR 7001) has been the primary spectrophotometric standard for many
years. The spectrum of Vega was investigated with high resolution
and high signal-to-noise ratio (S/N) (Gulliver et al. (1991),
Gulliver, Hill \& Adelman (1994), Takeda (2007)). Ultra-high S/N
($>$2500) reported in studies by Gulliver et al.(1991) allowed
them to recognize flattened shapes of cores of weak lines caused
by the fast and almost pole-on rotation of Vega (Gulliver et al.,
1994). The later studies (e.g., Hill, Gulliver \& Adelman (2004),
Peterson et al. (2006)) ascertained details of the nature of
Vega's rotation: the the star is seen almost pole-on, which causes
the flat-bottomed shape of profiles of weak lines.

  The chemical composition of Vega has well-known peculiarities:
the star demonstrates under-abundances in the majority of elements
and, has been regarded as a $\lambda$ Boo star (Adelman \&
Gulliver (1990), Lemke \& Venn (1996), Ilijic et al. (1998)). This
type of star is characterized by fast rotation, deficit of metals
(Baschek \& Slettebak, 1988) and slightly overabundant C, N and O
elements. The phenomenon of $\lambda$ Boo stars has been
attributed toe accretion from the circumstellar disk (Venn \&
Lambert, 1990). However, the extensive dust ring disc around Vega
(Aumann et al., 1984) has been observed in many wavelengths (e.g.,
Marsh et al., 2006). Waters et al. (1992) suggested that abundance
anomalies in Vega are caused by selective accretion: grains in the
circumstellar disc are accelerated outward by radiation pressure,
so that only the gaseous part of the disc is accreted.

\begin{figure*}
\caption{A comparison of (top) our atlas of Vega with (bottom)
atlas of Takeda et al.(2007). {\bf a.} An artefact in the atlas of
Takeda et al. (in the dotted rectangle); {\bf b.} The wavelength
range with telluric lines. Our atlas has no telluric lines. }
\epsfig{file=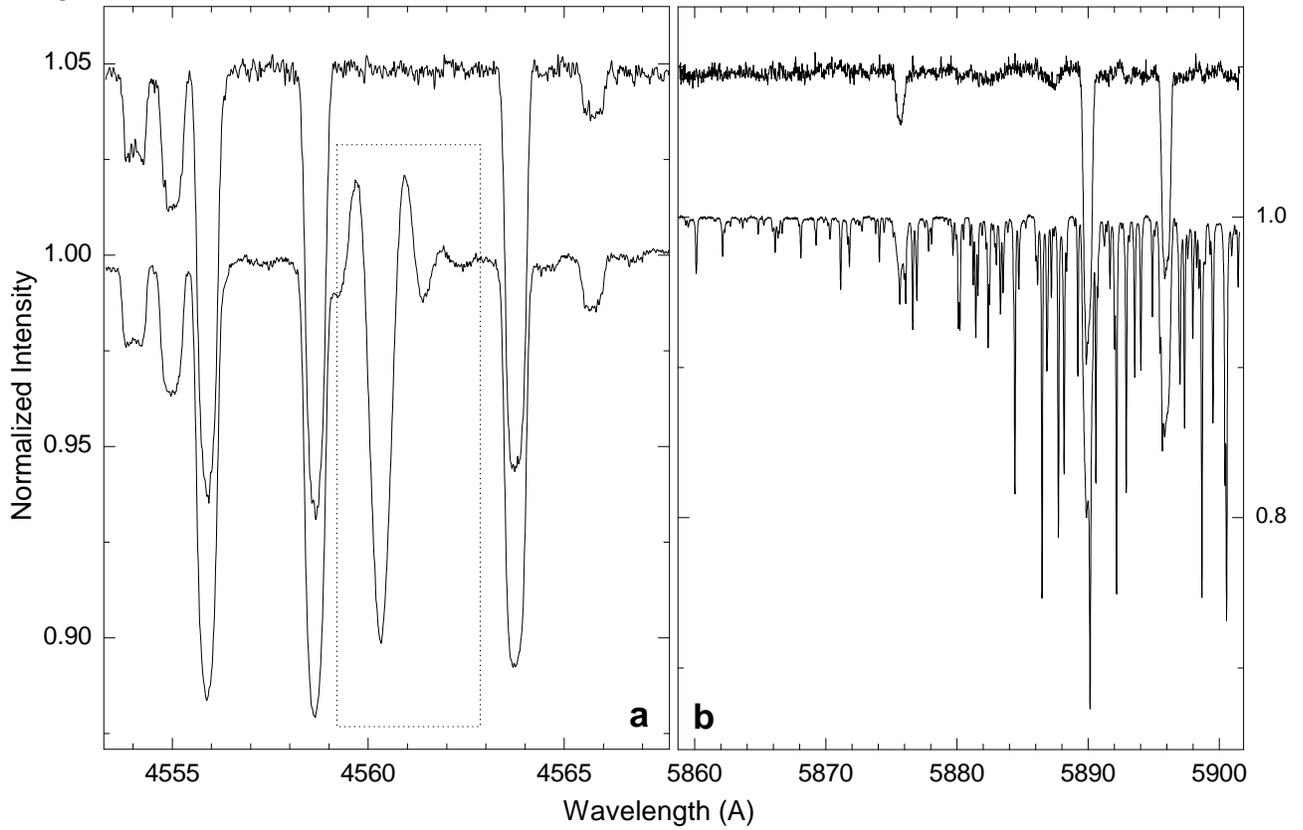, width=17cm} \label{takeda-we}
\end{figure*}

 Several spectral atlases of Vega  have been
published: an atlas in the region 2000 to 3187 \AA\ with
resolution of $\sim$0.1 \AA\ (Rogerson, 1989) and an atlas in the
4000--6350 \AA\ range with signal-to-noise ratio of about 300 (Qiu
et al., 1999). Gulliver \& Adelman (1990) published an atlas of
Vega with ultra-high signal-to-noise ($\sim$ 2500). The most
recent atlas was published by Takeda et al. (2007), who expanded
the wavelength range of the high quality spectrum, providing
detailed analysis of line profiles in the range 3900-8800 \AA.
However, the Takeda et al. atlas of Vega is not corrected for
telluric lines, although a spectrum of a fast rotating star is
given for a comparison. In some wavelength ranges, the Takeda et
al. atlas demonstrates features of instrumental or reduction
origin (see Fig. \ref{takeda-we}).

In this work, we present a high resolving power ($\sim$90,000) and
high S/N ratio ($\sim$700) spectrum of Vega covering the
wavelength region 3850 -- 6860 \AA.

The atlas is available in table form via the Internet.
\footnote{http://simbad.u-strasbg.fr}

\section{Observations and data reduction}

The spectra of Vega were recorded in 2006 during observing runs at
the Bohyunsan Observatory the aid of BOES, the fiber-fed echelle
spectrograph (Kim et al., 2007) attached to the 1.8-m telescope.
The spectrograph has 3 observational modes with three levels of
resolving power: 30,000, 45,000 and 90,000. The spectrograph has a
CCD camera equipped with a 4096$\times$2048 pixels matrix (pixel
size 15$\mu$m$\times$15$\mu$m) which allows us to cover in a
single exposure the wavelength range $\sim$3700~\AA\ --
$\sim$10000~\AA\ divided into 75-83 spectral orders without any
gaps up to $\sim$9500 \AA. Our spectra were recorded in the
highest resolving power mode (R = 90,000) corresponding to the
full width at the half maximum of the instrumental profile
$\sim$2.5 pixels.

The collected spectra were reduced using standard software packages
IRAF (Tody, 1993) and DECH (Galazutdinov, 1992). The latter was
used for final stages of processing: removal of telluric lines, merging of
spectra, wavelength scale correction, continuum normalization,
measurements of equivalent widths, etc. IRAF was used for bias/background
subtraction and one-dimensional spectra extraction.
Inter-order stray light in BOES spectra is less than 2 percent of
the intensities of the neighboring order (Kim et al., 2007). However,
the IRAF task APALL (we used it for the extraction of one-dimensional spectra from echelle images) does not provides any control of background
subtraction quality, therefore we used the APSCATTER task for careful background subtraction.
Traces of cosmic ray hits were removed during a stage of combining of individual spectra using an algorithm based on
a sigma-clipping method.

\begin{figure}
\caption{An example of removal of telluric lines. Top: a spectrum
after removal procedure. Middle: the spectrum before correction.
Bottom: a modified (stellar lines were eliminated) spectrum of a
divisor.} \epsfig{file=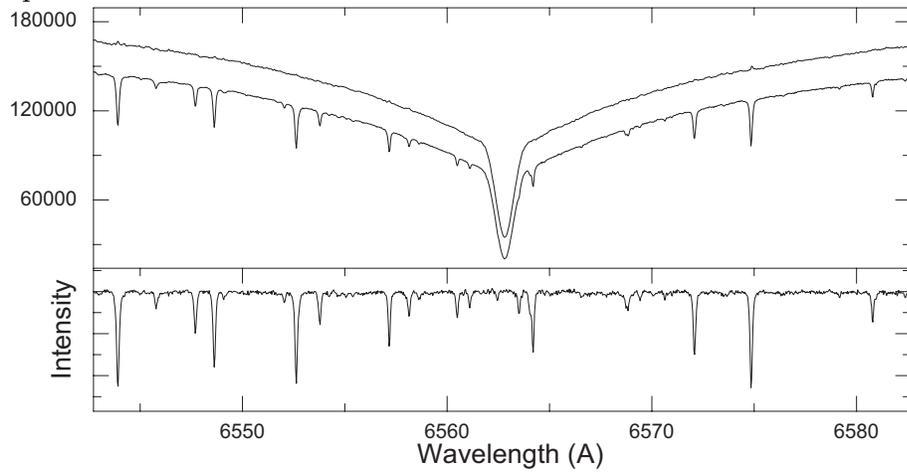, width=12cm} \label{telluric}
\end{figure}

\begin{figure}
\caption{An example of averaging of Vega spectra, rms deviation
curve shown in the lower portion. Note a positional accuracy of
wavelength shift of individual spectra. An example of instrumental
profile, constructed by averaging of 10 non-blended ThAr lines, is
given in the small box.} \epsfig{file=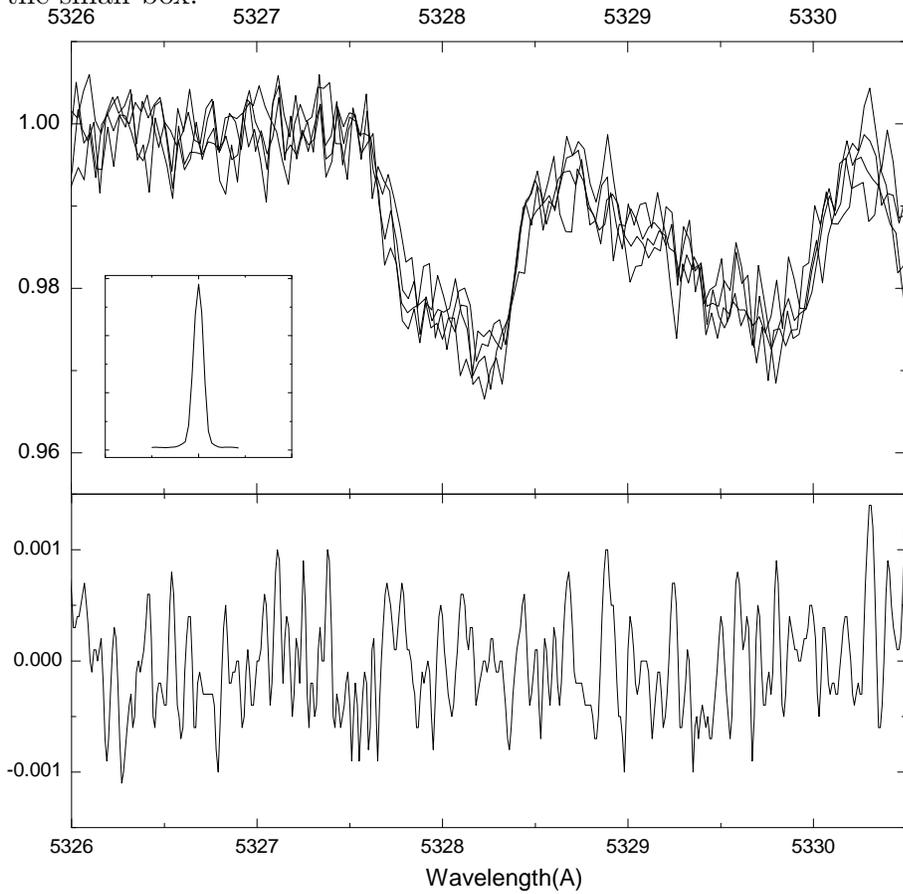, width=12cm}
\label{aver}
\end{figure}

The sequence of reduction stages after spectra extraction and wavelength calibration was as follows:
\begin{description}
 \item{\bf Averaging of spectra recorded during the same night.} Our spectra of Vega were acquired
 during 5 nights (Table \ref{table1}). All spectra of every single night were averaged with sigma-clipping filtering of cosmic-ray events.
 All subsequent procedures were performed with these 5 spectra.
 \item{{\bf Removal of telluric lines in each individual echelle order.}  To eliminate telluric lines, we used the spectrum
 of $\alpha$ Peg (a divisor) -- a fast rotating early spectral type star without significant reddening, observed during
 the same nights as the Vega observations. Before the usage of the divisor, all stellar lines existing in its spectrum
 were removed using the pseudo-continuum (the curve, passing over all stellar lines).
 An example of the effect of removal of telluric lines  is presented in Fig. \ref{telluric}}
 \item {{\bf Shift of the wavelength scale of individual spectra to the laboratory system.} This procedure is important
 for the next step, spectra averaging. We measured radial velocity of H$_{\alpha}$ in each single spectrum and, then by using this
 velocity value we calculated and applied the wavelength correction for each pixel of the spectrum. The H$_{\alpha}$ line profile
 is quite symmetric and occupies many pixels (in comparison to weaker lines of, e.g. Fe{\sc
 i}),which facilitates finding the center of gravity of the line, based on the method of matching direct and mirrored profiles, which we used.
 The quality of averaging of the spectra is demonstrated in Fig. \ref{aver}.}
 \item {{\bf Averaging and merging of all spectra.} All spectral orders were merged. Then, all individual spectra were averaged.}
 \item {{\bf Continuum normalization.} The final spectrum was normalized by the fiducial continuum.  However, in the region of very broad hydrogen lines
 the continuum level was adjusted with the help of the theoretical calculations (the synthetic spectrum) - i.e.. the resulting hydrogen line profiles are not
 completely observational. }
\end{description}

\begin{figure}
\caption{The iron lines around 4530 \AA} \epsfig{file=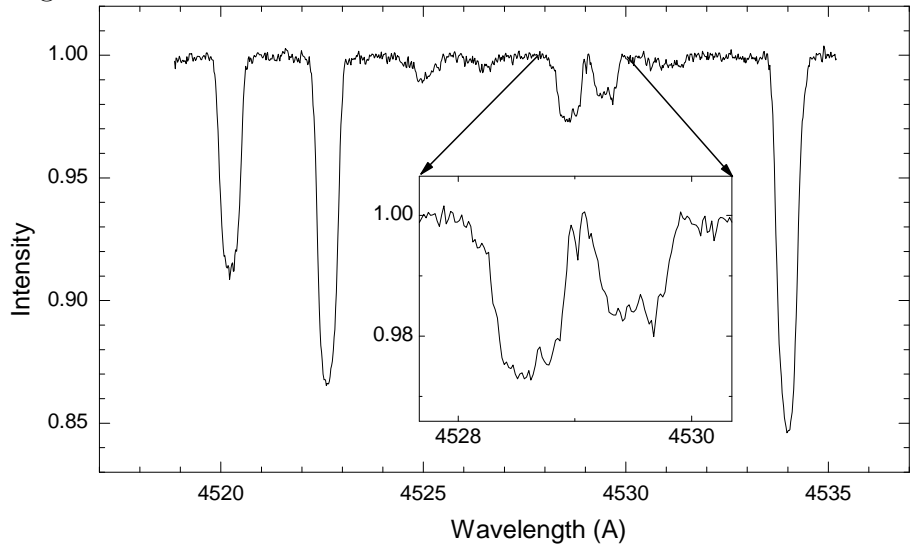,
width=12cm} \label{4530}
\end{figure}

We follow the tradition of showing iron lines around 4530 \AA,
started by Gulliver et al. in 1991 (see Fig. \ref{4530}). However,
our profiles are not smooth like Gulliver's  but rather
demonstrate some pattern, which is also the case in the spectrum
of Takeda et al.(2007; see their Fig.5).

\section{Summary}

 In summary, we present a high resolution and high signal-to-noise
ratio atlas of Vega. Data are downloadable in digital form. We
hope that the atlas provides useful information for future
studies.

\section{Acknowledgements}

GAG and IH acknowledge support by KICOS through the grant No. 07-179.

\clearpage

\begin{table}
\caption{List of Vega spectra. }
\label{table1}
\centering
\begin{tabular}{c c c c  c c}
\hline\hline
     Date & Number     & HJD$^\mathrm{a}$ &                & S/N            & \\
          \cline{4-6}
          & of spectra &                  & $\sim$4000 \AA & $\sim$5500 \AA & $\sim$6500 \AA \\
\hline
2006 Mar 30  &  10   &  2453824.2701    &  195  &  422  &  406  \\
2006 Mar 31  &  28   &  2453825.3104    &  235  &  394  &  434  \\
2006 Apr 16  &  21   &  2453841.2743    &  237  &  387  &  442  \\
2006 Jun 11  &  15   &  2453897.1451    &  203  &  426  &  286  \\
2006 Oct 03  &  10   &  2454011.9792    &  355  &  478  &  499  \\
\hline
\end{tabular}
\begin{list}{}{}
\item[$^{\mathrm{a}}$] Heliocentric Julian date of middle of exposure
\end{list}
\end{table}

\end{document}